\documentclass[reprint,superscriptaddress,a4paper,aps,longbibliography]{revtex4-1}
\usepackage{graphicx}
\usepackage{amsmath}

\begin{document}

\title{Slow spin tunneling in the paramagnetic phase of the pyrochlore Nd$_2$Sn$_2$O$_7$} 

\author{P. Dalmas~de~R\'eotier}
\affiliation{Universit\'e Grenoble Alpes, INAC-PHELIQS, F-38000 Grenoble, France}
\affiliation{CEA, INAC-PHELIQS, F-38000 Grenoble, France}
\author{A. Yaouanc}
\affiliation{Universit\'e Grenoble Alpes, INAC-PHELIQS, F-38000 Grenoble, France}
\affiliation{CEA, INAC-PHELIQS, F-38000 Grenoble, France}
\author{A. Maisuradze}
\affiliation{Department of Physics, Tbilisi State University, Chavchavadze 3, GE-0128
  Tbilisi, Georgia}
\author{A. Bertin}
\affiliation{Universit\'e Grenoble Alpes, INAC-PHELIQS, F-38000 Grenoble, France}
\affiliation{CEA, INAC-PHELIQS, F-38000 Grenoble, France}
\author{P. J. Baker}
\affiliation{ISIS Facility, STFC Rutherford Appleton Laboratory, Chilton, Didcot, OX11 0QX, UK}
\author{A. D. Hillier}
\affiliation{ISIS Facility, STFC Rutherford Appleton Laboratory, Chilton, Didcot, OX11 0QX, UK}
\author{A.~Forget}
\affiliation{SPEC, CEA, CNRS, Universit\'e Paris-Saclay, CEA Saclay, 91191 Gif-sur-Yvette Cedex, France}

\date{\today}

\begin{abstract}

  The insulating pyrochlore compound Nd$_2$Sn$_2$O$_7$ has been shown to undergo a second order magnetic phase transition at $T_{\rm c}  \approx 0.91 $~K to a noncoplanar all-in--all-out magnetic structure of the Nd$^{3+}$ magnetic moments. An anomalously slow paramagnetic spin dynamics has been evidenced from neutron backscattering and muon spin relaxation ($\mu$SR). In the case of $\mu$SR this has been revealed through the strong effect of a 50~mT longitudinal field on the spin-lattice relaxation rate. Here, motivated by a recent successful work performed for Yb$_2$Ti$_2$O$_7$ and Yb$_2$Sn$_2$O$_7$, analyzing the shape of the $\mu$SR longitudinal polarization function, we substantiate the existence of extremely slow paramagnetic spin dynamics in the microsecond time range for Nd$_2$Sn$_2$O$_7$. Between 1.7 and 7~K, this time scale is temperature independent. This suggests a double spin-flip tunneling relaxation mechanism to be at play, probably involving spin substructures such as tetrahedra. Unexpectedly, the standard deviation of the field distribution at the muon site increases as the system is cooled. This exotic spin dynamics is in sharp contrast with the dynamics above 100~K which is driven by the Orbach relaxation mechanism involving single Nd$^{3+}$ magnetic moments.

\end{abstract}

\maketitle

\section{Introduction}
Exotic magnetic fluctuations and correlations are expected to be observed for geometrically frustrated magnetic materials \cite{Moessner06,Gardner10,Balents10,Lacroix11,Gingras14}. Probably, the best documented experimental example is given by the ordered state of the pyrochlore insulator compound Tb$_2$Sn$_2$O$_7$ for which signatures of unconventional fluctuations have been found from muon spin relaxation ($\mu$SR), different types of neutron scattering techniques and specific heat data \cite{Dalmas06,Bert06,Chapuis07,Rule07,Mirebeau08,Rule09b,Bonville10,Dalmas16a}. The existence of unexpected short-range correlations has been pointed out for the ordered state of the pyrochlore insulators Er$_2$Ti$_2$O$_7$ \cite{Ruff08}, Yb$_2$Ti$_2$O$_7$ \cite{Bonville03a,Ross11a,Maisuradze15}, Yb$_2$Sn$_2$O$_7$ \cite{Maisuradze15} and the triangular system  La$_2$Ca$_2$MnO$_7$ \cite{Dalmas15a}.  Exotic fluctuations have been discovered for cooperative paramagnets such as the spin-1/2 kagome lattice herbertsmithite ZnCu$_3$(OH)$_6$Cl$_2$ \cite{Fu15}, the pyrochlore insulator Tb$_2$Ti$_2$O$_7$ \cite{Ueland06} and the triangular system NiGa$_2$S$_4$ \cite{Yaouanc08,Nambu15}. A large range of fluctuation rates is usually observed, as documented in Ref.~\onlinecite{Nambu15}. However, the number of compounds which display magnetic ordering at low temperature and unconventional paramagnetic fluctuations is still restricted. Recently, the normal spinel CdHo$_2$S$_4$ \cite{Yaouanc15}, and the pyrochlore insulators Yb$_2$Ti$_2$O$_7$, Yb$_2$Sn$_2$O$_7$, Nd$_2$Sn$_2$O$_7$, Nd$_2$Zr$_2$O$_7$, and Er$_2$Ti$_2$O$_7$  have been shown to belong to this family of compounds \cite{Maisuradze15,Bertin15,Xu16,Orendac16}.
 
Anomalously slow paramagnetic fluctuations with a correlation time roughly in the 100~ns time range have been reported for the cubic pyrochlore stannate Nd$_2$Sn$_2$O$_7$ \cite{Bertin15}. They were unraveled through the study of the influence of an external magnetic field of 50~mT on the $\mu$SR spin-lattice relaxation rate. It has been shown recently that the study of the shape of the $\mu$SR longitudinal polarization function can be a very effective method to detect anomalously slow paramagnetic fluctuations \cite{Maisuradze15}. Motivated by this result, we have performed a low field $\mu$SR study of the paramagnetic state of Nd$_2$Sn$_2$O$_7$ to further characterize its spin dynamics. Here we report on this detailed study. 

Nd$_2$Sn$_2$O$_7$ crystallizes in the pyrochlore structure in which the rare earth ions form a network of corner sharing regular tetrahedra. It exhibits a magnetic phase transition at $T_{\rm c} \approx0.91$~K to a so-called all-in--all-out magnetic structure: the magnetic moment of an ion is collinear to the direction linking the corner at which the ion sits to the tetrahedron center and all four moments of a given tetrahedron point either inwards or outwards. A $\mu$SR spontaneous field has been observed, consistent with the lack of the divergence-free part of the Helmholtz decomposition of the magnetic-moment field for such a magnetic structure \cite{Brooks14}. In this frame, the long-range order is associated with the divergence-full component of the field. Persistent spin dynamics below $T_{\rm c}$ and anomalously slow paramagnetic fluctuations up to $\approx 30 \, T_{\rm c}$ have been detected. 

We shall first describe the experimental and data analysis methods in Sec.~\ref{Experimental}. Our experimental results will be presented in Sec.~\ref{Results} and discussed in Sec.~\ref{Discussion}. Finally, in Sec.~\ref{Conclusions} conclusions are gathered.

\section{Experimental and data analysis}
\label{Experimental}

The measurements were performed with a powder sample previously used for a $\mu$SR work \cite{Bertin15}. The asymmetry spectra were recorded at the MuSR and EMU spectrometers of the ISIS facility (Rutherford Appleton Laboratory). The sample was mixed with a small fraction of General Electric varnish and deposited to a silver plate.

The longitudinal-field geometry was adopted, for which an external field ${\bf B}_{\rm ext}$, if any, is set along the initial muon beam polarization ${\bf S}_\mu$ \cite{Yaouanc11}. By definition, ${\bf B}_{\rm ext}$ is applied along the $Z$ axis of the laboratory reference frame.

The measured longitudinal-field (LF) asymmetry spectra will be analyzed with the two-component model
\begin{equation}
 A^{\rm LF}(t) = A^{\rm LF}_0 \left[ (1-F_{\rm bg})P_{Z, {\rm NSO}}(t) + F_{\rm bg} P_{Z,{\rm bg}} (t)\right],
 \label{eq:AsyModel:LF}
\end{equation}
where $A^{\rm LF}_0$ is the initial $\mu$SR asymmetry. The longitudinal polarization function $P_{Z, {\rm NSO}}(t)$ describes the evolution of the $Z$ component of the muon spin ensemble \cite{Yaouanc11}. $F_{\rm bg}$ is the fraction of muons stopped outside of the sample, mainly in the silver backing plate and possibly in the cryostat window or walls. A weakly damped background function is required to account for the spectra at low fields. We find that $P_{Z,{\rm bg}} (t)$ can be taken as the so-called static Kubo-Toyabe (KT) function \cite{Hayano79}. 

The simplest possible model for $P_{Z, {\rm NSO}}(t)$ is also the KT model. It depends on the mean field $B_{\rm LF}$ and the standard deviation $\Delta$ of the field distribution at the muon site assumed to be Gaussian and isotropic. Spin dynamics is described with the strong collision model \cite{Kehr78} which requires the introduction of a field fluctuation frequency, $\nu_{\rm c} = 1/\tau_{\rm c}$. Here $\tau_{\rm c}$ is the characteristic decay time of the correlation function of the field experienced by the muons. We have $P_{Z, {\rm NSO}}(t) = P_{Z, {\rm KT}}(t,B_{\rm LF},\Delta,\nu_{\rm c})$ \cite{Yaouanc11}.

\section{Experimental results}
\label{Results}

\begin{figure}
\includegraphics[width=0.49\linewidth]{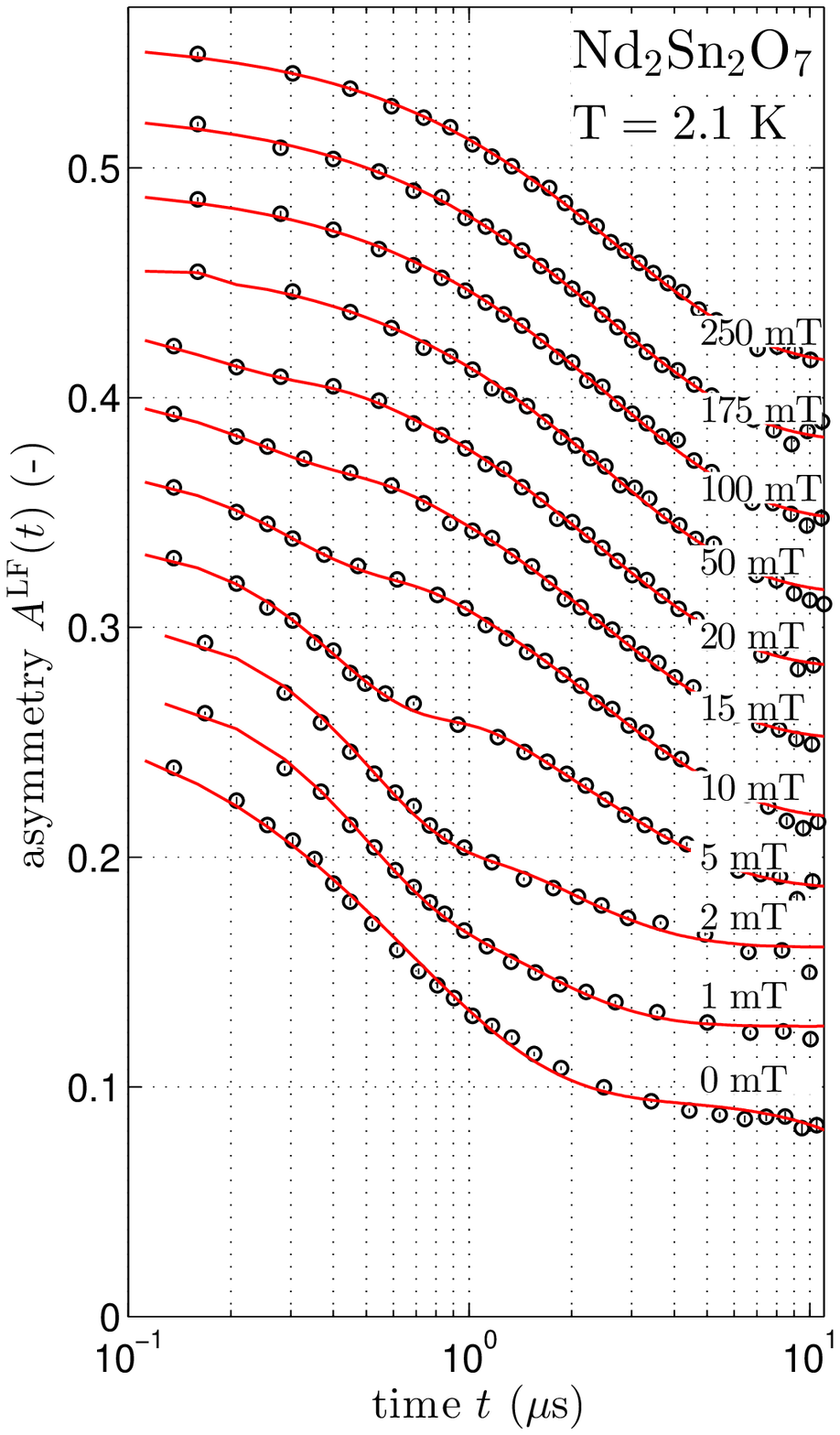}
\includegraphics[width=0.49\linewidth]{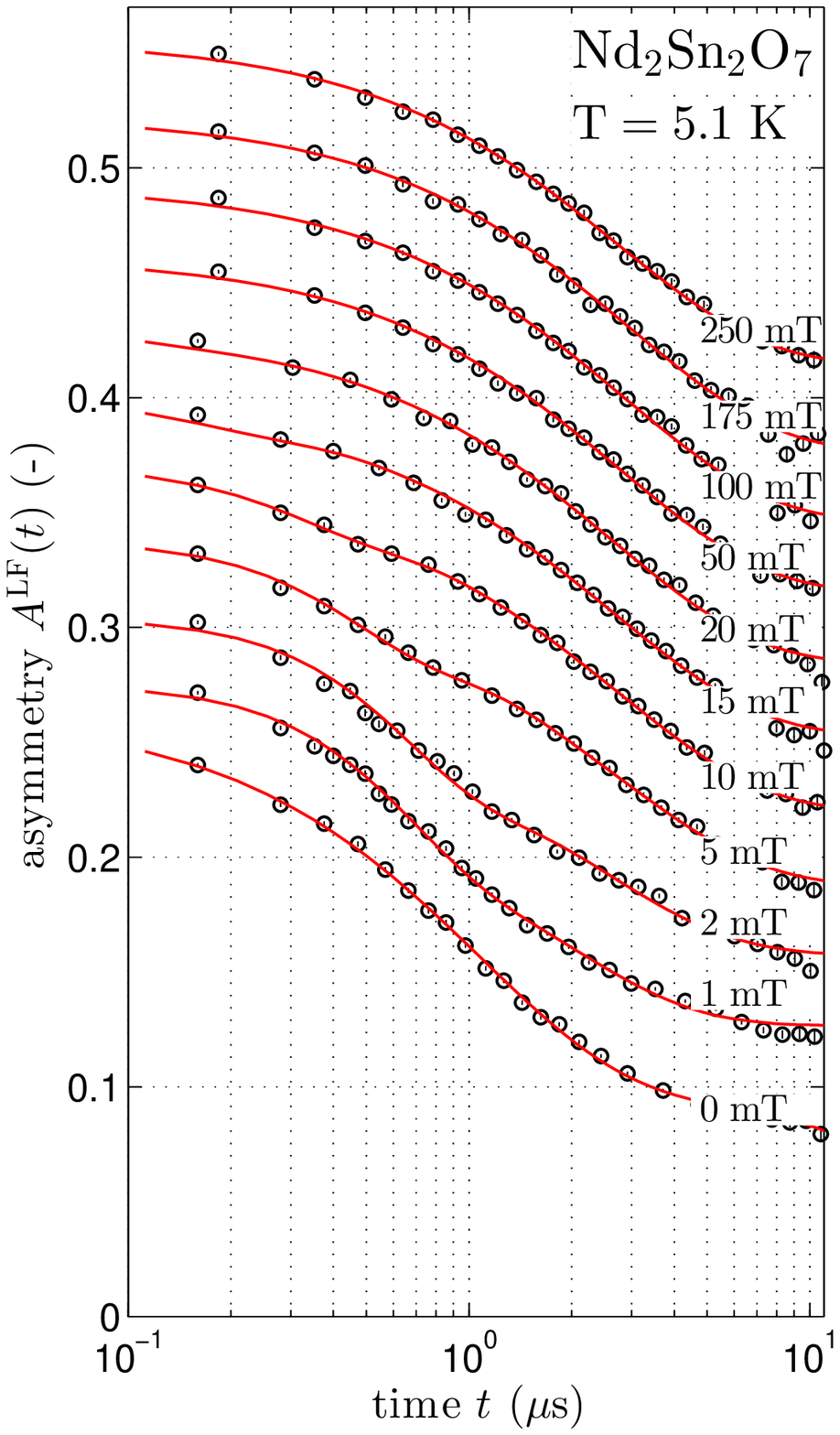}
\includegraphics[width=0.49\linewidth]{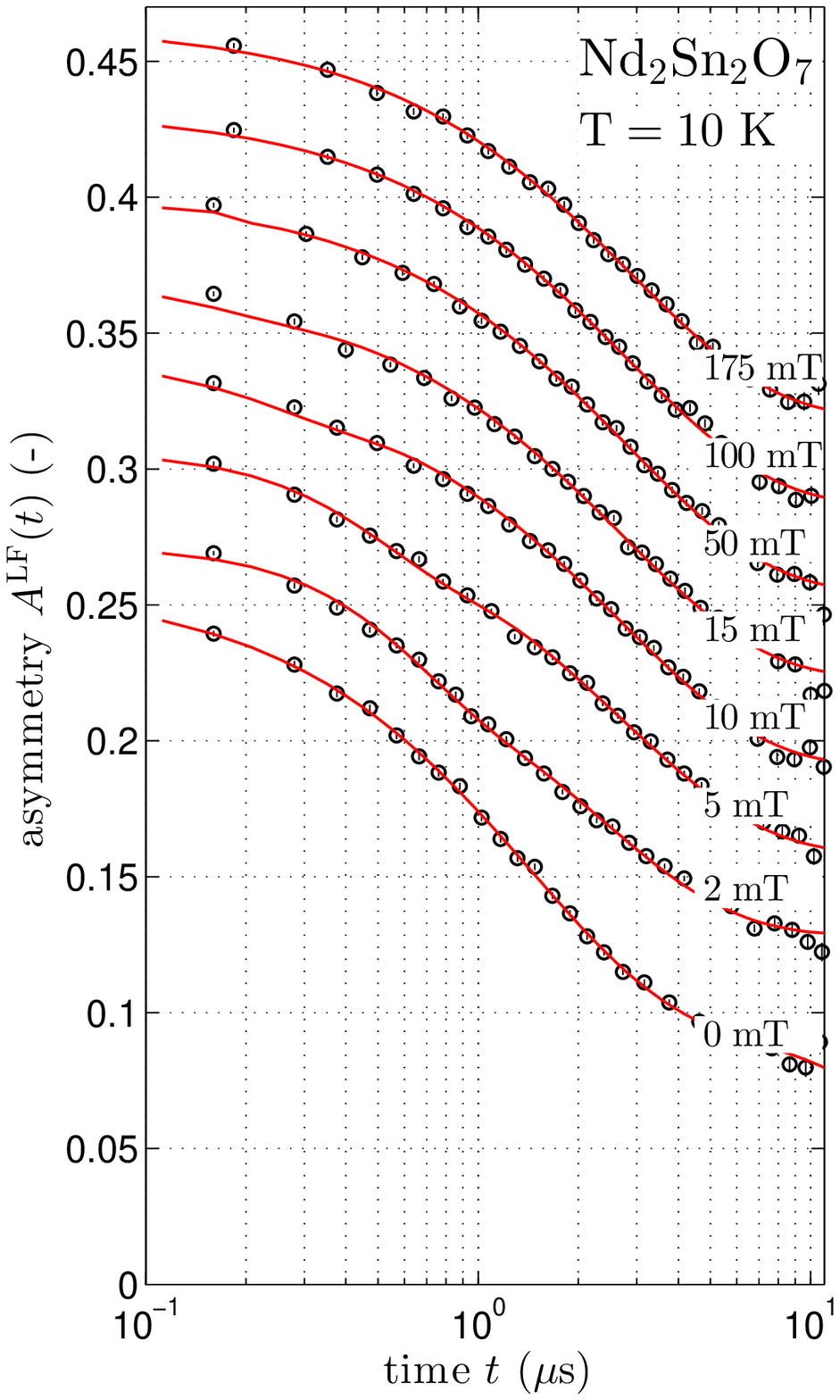}
\includegraphics[width=0.49\linewidth]{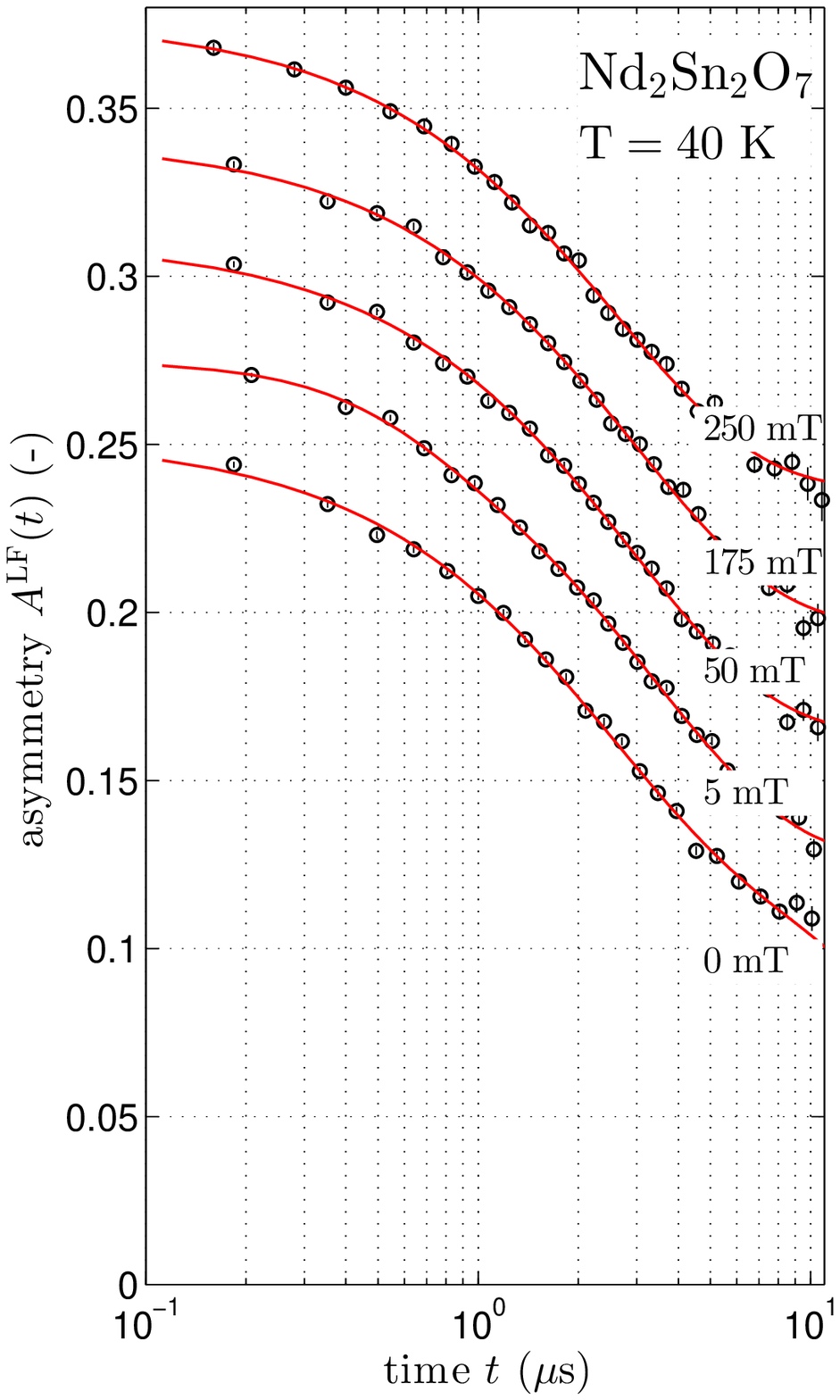}
\caption{(color online). Asymmetry spectra recorded in the paramagnetic phase of a Nd$_2$Sn$_2$O$_7$ powder sample at four temperatures in various longitudinal fields. The spectra were recorded at MuSR. The experimental points are represented by black circles. The solid lines result from fits with the conventional dynamical KT model as explained in the main text. The spectra for consecutive external field values (indicated next to each spectrum) are vertically shifted by 0.03 unit for better visualization.  The statistical uncertainties are indicated by vertical bars, the size of which is much less than the symbol size except near 10~$\mu$s. Note that the spectra are displayed with a logarithmic time scale.
}
\label{Nd2Ti2O7_para_spectra_versus_field}
\end{figure}

Examples of asymmetry spectra recorded with the MuSR spectrometer at four temperatures and the related fitting curves are presented in Fig.~\ref{Nd2Ti2O7_para_spectra_versus_field}. 
In contrast to Yb$_2$Sn$_2$O$_7$ and Yb$_2$Ti$_2$O$_7$ \cite{Maisuradze15}, the asymmetry monotonically decays with time at all fields for the four temperatures. So the signature of quasi-static fluctuations is not so obvious graphically as it was for the ytterbium compounds. However, a careful inspection of the spectra reveals that some of them depart from an exponential or stretched-exponential decay, i.e.\ the spin dynamics they probe is quasi-static. For instance, consider the spectra at 2.1~K. A break in the slope around $0.5\ \mu{\rm s}$ is found for the spectra with $B_{\rm ext} = 5$~mT and nearby values. 

The curves in Fig.~\ref{Nd2Ti2O7_para_spectra_versus_field} result from fits with Eq.~\ref{eq:AsyModel:LF}. The fits were performed with common values for $F_{\rm bg}$ and the field width $\delta_{\rm bg}$ associated with $P_{Z,{\rm bg}} (t)$. Parameters $\Delta$ and $\nu_c$ tend to increase with $B_{\rm ext}$, in a manner reminiscent to what was observed for Yb$_2$Ti$_2$O$_7$ and Yb$_2$Sn$_2$O$_7$ \cite{Maisuradze15}. The overall quality of the fits is extremely good, with the exception of the low field spectra at the lowest temperature. We will return to this point at the end of Sect.~\ref{Results}.

\begin{figure}
\centering
\includegraphics[width=0.60\linewidth]{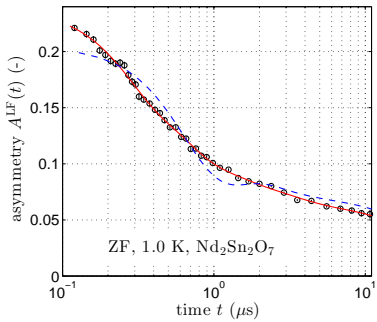}
\caption{(color online). 
Short-range correlations or two muon sites at low temperature in the paramagnetic phase of a Nd$_2$Sn$_2$O$_7$ powder sample. We display a ZF asymmetry spectrum recorded at 1.0~K with the MuSR spectrometer. The experimental points are represented by black circles and the corresponding fitting curve with short-range correlations taken into account and two muon sites as a red solid line. The blue dashed line results from the best fit with a single dynamical KT function (Eq.~\ref{eq:AsyModel:LF}).}
\label{Nd2Ti2O7_para_spectra_lowT}
\end{figure}

To ascertain the relevance of the observed deviations from the dynamical KT model as the sample is cooled down, we have extended the zero-field measurements to two lower temperatures. An example is presented  in Fig.~\ref{Nd2Ti2O7_para_spectra_lowT}.
The dashed line clearly shows that the conventional Kubo-Toyabe function no longer provides an adequate fit to the data. Instead, the full line provides a remarkable fit. Here the dynamical KT function is replaced by an equal-weight sum of two polarization functions for which the KT Gaussian field distribution is extended to account for short-range correlations (SRC) \cite{Yaouanc13a,Maisuradze15}. Alternative fits with either a weighted sum of two dynamical KT functions or a single SRC function still provide reasonable descriptions of the data (not shown). We cannot select the suitable model with the available information.  However, the value of $\nu_{\rm c}$ extracted from the three models is the same within statistical uncertainties. 

\begin{figure}
\centering
\includegraphics[width=0.60\linewidth]{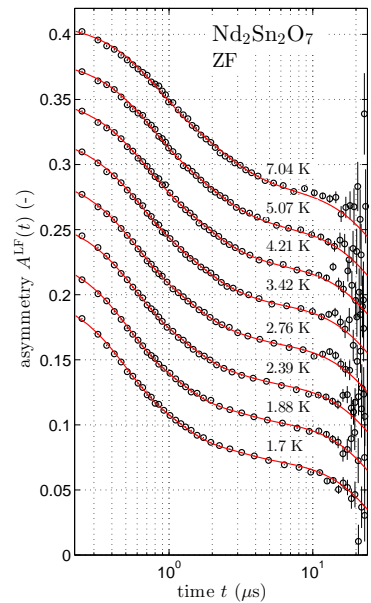}
\caption{(color online). ZF asymmetry spectra recorded at low temperature with the EMU spectrometer. The experimental points are represented by black circles. The solid lines result from fits explained in the main text. The spectra for consecutive temperature values (indicated next to each spectrum) are vertically shifted by 0.03 unit for better visualization.}
\label{Nd2Sn2O7_ZF_EMU}
\end{figure}

\begin{figure}
\centering
\includegraphics[width=0.60\linewidth]{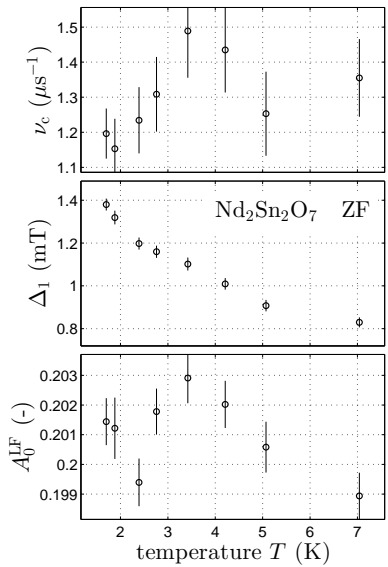}
\caption{Thermal dependence of  $\nu_{\rm c}$, $\Delta_1$, and $A^{\rm LF}_0$ extracted from the analysis of ZF spectra recorded at the EMU spectrometer.}
\label{Nd2Sn2O7_para_versus_temp}
\end{figure}

For a better determination of the observed deviations from the conventional KT model seen in ZF, we have performed ZF measurements with high statistics --- about $80 \times 10^6$ decay positrons recorded per spectrum --- at the EMU spectrometer. The data were recorded at 7~K and below where the spectra substantially deviate from an exponential-like relaxation. The results are shown in Fig.~\ref{Nd2Sn2O7_ZF_EMU}.
A successful combined fit of the eight spectra has been done with a weighted sum of two dynamical KT functions, i.e.\ $P_{Z, {\rm NSO}}(t) = \sum_{i=1}^{2} F_i\,P_{Z,{\rm KT}}(t, B_{\rm LF}=0,\Delta_i,\nu_c)$ with $F_1+F_2=1$. This model requires a minimum number of free parameters. For the background we have $F_{\rm bg} = 0.366 \, (1)$ and $\delta_{\rm bg} = 0.041 \, (1)$~mT. These values are in the expected range for the background signal.
The relative weight of the first component is $F_1 = 0.62 \, (1)$. We have assumed $\nu_{\rm c}$ to be the same for the two relaxation functions and a common ratio for the two standard deviations for the eight spectra. The value for the ratio $\Delta_2/\Delta_1 = 2.11 \, (2)$ was set from a prior fit with independent values of $\Delta_1$ and $\Delta_2$. The thermal dependences of the remaining free parameters, namely $\nu_{\rm c}$, $\Delta_1$ and $A^{\rm LF}_0$, are displayed in Fig.~\ref{Nd2Sn2O7_para_versus_temp}.
Within experimental uncertainties, $A^{\rm LF}_0$ is temperature independent, as it should. We observe a pronounced increase of $\Delta_1$ as the sample is cooled down, and this parameter is much smaller than one would have expected for the standard deviation of a field distribution arising from Nd$^{3+}$ magnetic moments. The spin dynamics characterized by $\nu_{\rm c}$ is anomalously slow. Based on the Curie-Weiss temperature $\Theta_{\rm CW} = -0.32\,(1)$~K \cite{Bertin15} of Nd$_2$Sn$_2$O$_7$, we would have expected $\nu_{\rm c}$ of order $k_{\rm B}|\Theta_{\rm CW}|/\hbar \approx 4\times 10^{10}$~s$^{-1}$, i.e. more than four orders of magnitude larger than observed. Here $k_{\rm B}$ and $\hbar$ are the Boltzmann and Dirac constants, respectively. The extremely small $\nu_{\rm c}$ value previously roughly inferred from the anomalous field dependence of the $\mu$SR relaxation rate \cite{Bertin15} is therefore confirmed. The temperature independence of $\nu_{\rm c}$ (Fig.~\ref{Nd2Sn2O7_para_versus_temp}) is another remarkable feature. On the approach of a phase transition, but still outside the critical regime, it is expected that $\nu_c(T) \propto (T-T_{\rm c})$, $T_{\rm c}$ being a Curie or N\'eel temperature \cite{Moriya62,deGennes60}, at odds from our result.

\begin{figure}
\centering
\includegraphics[width=0.60\linewidth]{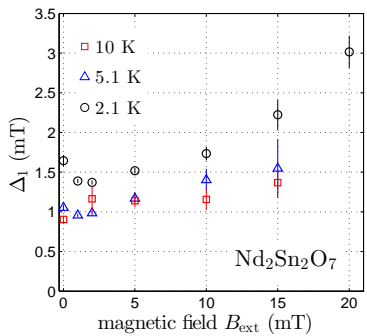}
\caption{(color online). Field dependence of $\Delta_1$ obtained from the two dynamical KT function model applied to the data shown in Fig.~\ref{Nd2Ti2O7_para_spectra_versus_field}. When the spectral shape is too close to an exponential, the fits do not allow for an accurate determination of the parameters, which are therefore not displayed. }
\label{Nd2Sn2O7_2KT_Delta}
\end{figure}

The two-component model just used for the high statistics spectra allows us to satisfactorily fit the whole set of data shown in Fig.~\ref{Nd2Ti2O7_para_spectra_versus_field}, including the low field spectra at 2.1~K. Taking the example of the zero-field spectrum recorded at 2.1~K, the confidence parameter $\chi^2$ changes from 1.33 to 0.97 for the one and two-component model, respectively. Figure~\ref{Nd2Sn2O7_2KT_Delta} displays $\Delta_1(B_{\rm ext})$. The field width $\Delta_1$ increases with $B_{\rm ext}$, especially at low temperature similarly to what was observed in the ytterbium titanate and stannate \cite{Maisuradze15}.

Going back to the thermal dependence of the high statistics spectra, the thermal behavior of the two parameters characterizing the muon spin relaxation is exotic. The signature of the approach to the second-order phase transition is seen in $\Delta_1 (T)$ rather than in $\nu_{\rm c} (T)$. This is further discussed in the next section, together with the $\Delta_1$ and $\nu_{\rm c}$ values. Before that, it is interesting to notice that these results are partly reminiscent of those recently obtained for Yb$_2$Ti$_2$O$_7$ and Yb$_2$Sn$_2$O$_7$ \cite{Maisuradze15}. These compounds display splayed ferromagnetic ground states \cite{Yaouanc13,Yaouanc16}, completely different from the antiferromagnetic ground state of  Nd$_2$Sn$_2$O$_7$. In addition, they are characterized by first order magnetic phase transitions, again in contrast to  Nd$_2$Sn$_2$O$_7$. This suggests that the exotic paramagnetic fluctuations we observe are independent of the type of magnetic ordering at low temperature. They rather seem to reflect the frustrated magnetic interactions present in these rare-earth pyrochlore compounds.

\section{Discussion}
\label{Discussion}

The existence of two relaxation channels has recently been inferred from ac susceptibility measurements at low temperature for Er$_2$Ti$_2$O$_7$ \cite{Orendac16}: an Orbach relaxation with a remarkable small activation energy and an attempt time much longer than expected, and a temperature independent relaxation. Here we shall discuss our results in relation to the interpretation of the Er$_2$Ti$_2$O$_7$ susceptibility data. We shall first examine the possible reason for the absence of the first relaxation mechanism in the $\mu$SR data and consider the $\Delta_1$ and $\nu_{\rm c}$ parameters in turn.

\begin{figure}
\centering
\includegraphics[width=0.80\linewidth]{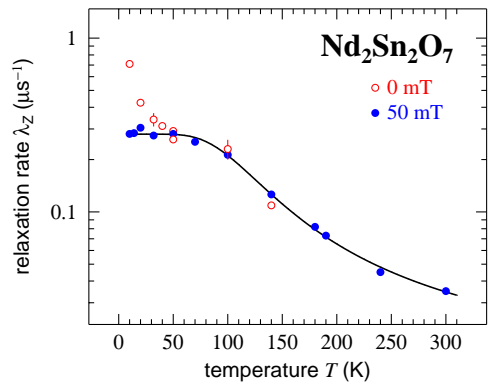}
\caption{(color online). Thermal dependence of the $\mu$SR relaxation rate $\lambda_Z$ measured in 0 or 50~mT longitudinal field in Nd$_2$Sn$_2$O$_7$ for temperatures larger than 10~K. The data are taken from Ref.~\cite{Bertin15}. The full line displays $\lambda_Z(T)$ for an Orbach relaxation mechanism: $\lambda_Z^{-1}(T) = A_{\rm O} + B_{\rm O}\exp(-\Delta_{\rm CEF}/k_{\rm B}T)$ \cite{Dalmas04a}. The fit to the data gives $A_{\rm O}^{-1}= 0.28\,(1)~\mu$s$^{-1}$ and $B_{\rm O}$ = 118\,(6)~$\mu$s, with $\Delta_{\rm CEF}$ = 39.8~meV. The difference below $\approx 30$~K between the data recorded in 0 and 50~mT field has been discussed in Ref.~\cite{Bertin15}. }
\label{Orbach}
\end{figure}

The traditional Orbach process involves two real phonons with an excited crystal-electric-field (CEF) level as intermediate \cite{Orbach61}. The spin-lattice relaxation measured by $\mu$SR above $\approx 100$~K in Nd$_2$Sn$_2$O$_7$ is well accounted for with an activation energy $\Delta_{\rm CEF}$ = 39.8~meV, i.e.\ the energy of the third excited CEF energy level \cite{Bertin15a}; see Fig.~\ref{Orbach}. Similarly, an Orbach process explains the spin dynamics of the geometrically frustrated garnet Yb$_3$Ga$_5$O$_{12}$ above 100~K \cite{Dalmas03} and also for Yb$_2$Ti$_2$O$_7$ \cite{Dalmas04a} and Nd$_2$Zr$_2$O$_7$ \cite{Xu16}. The scheme of the CEF levels  of Nd$^{3+}$ ions in Nd$_2$Sn$_2$O$_7$ cannot explain our experimental findings at low temperature since the first excited CEF doublet is located at 26~meV above the ground state \cite{Bertin15a}. 

The same energy misfit exists for Er$_2$Ti$_2$O$_7$. For this compound it has been advocated that the four spins sitting at the corner of a tetrahedron rather than single spins are the building blocks which generate the relaxation measured with ac susceptibility \cite{Orendac16}. This picture explains qualitatively the measured activation energy, and therefore justifies the existence of an Orbach relaxation, although with an attempt time much longer than expected. Its renormalization has been attributed to the bottleneck effect \cite{Abragam70}. To understand the experimental data for Er$_2$Ti$_2$O$_7$ a second temperature independent relaxation had to be added. This thermal behavior suggests a spin tunneling relaxation mechanism. The energy level scheme for a tetrahedron might support it through double spin-flip relaxation processes \cite{Bloembergen59}.

{\it A priori} the same picture should apply to  Nd$_2$Sn$_2$O$_7$. However, while the temperature independent relaxation channel has been found, an Orbach relaxation has not been detected. Recall that when two relaxation mechanisms with vastly different fluctuation frequencies  are present, only the mechanism falling into the technique time window should drive the relaxation. An Orbach relaxation with an attempt time in the range of a second and an activation energy of $\approx 1 $~meV as for Er$_2$Ti$_2$O$_7$ \cite{Orendac16} cannot contribute to the $\mu$SR relaxation. We propose this is also the case for Nd$_2$Sn$_2$O$_7$.

So we are left with the discussion of the characteristic parameters $\Delta_1$ and $\nu_{\rm c}$ for the remaining relaxation channel. The standard deviation $\Delta_1$ is much smaller than expected if there were no correlation between the spins. Indeed, following the methodology developed by van Vleck in the absence of correlations \cite{vanVleck48}, a field width of the order of a few hundred milliteslas is computed for different possible muon sites. Going further than this qualitative remark would require to model the static wavevector-dependent susceptibility and to know the muon site \cite{Dalmas04}. Concerning the fluctuation rate, we measure  $\nu_{\rm c} \simeq 1~\mu {\rm s}^{-1}$ while for Er$_2$Ti$_2$O$_7$ the value derived from ac susceptibility is approximately four orders of magnitude smaller. Since there is no detailed prediction available for the double spin-flip relaxation process, we cannot give a quantitative discussion and compare the results for Nd$_2$Sn$_2$O$_7$  and Er$_2$Ti$_2$O$_7$ in terms of their basic physical properties. 

\section{Conclusions}
\label{Conclusions}

The reported $\mu$SR measurements in the low temperature range of the paramagnetic phase of Nd$_2$Sn$_2$O$_7$ point to the presence of anomalously slow spin tunneling dynamics. The magnetic fluctuation frequency  is unexpectedly almost temperature independent within a relatively broad temperature range, while the standard field distribution at the muon site is anomalously small and reduces with increasing temperature.

These results suggest to focus attention on coupled spins within tetrahedra rather than to ionic spins. This is in sharp contrast to the dynamics above 100~K which only involves ionic spins. 

Similarly, tetrahedra spin structures should also influence the low temperature properties of other pyrochlore compounds such as Yb$_2$Sn$_2$O$_7$ and Yb$_2$Ti$_2$O$_7$ for which a detailed study has recently been published \cite{Maisuradze15}. Interestingly, the physical properties of tetrahedra of spins may depend on characteristics of the fictitious spins 1/2. The time reversal symmetry property of the ground state doublets, i.e.\ Kramers vs non Kramers, and their spatial symmetry, i.e.\ vector-like or multipole-like, are expected to play key roles \cite{Curnoe07a,Onoda10,Onoda11b,Onoda11b,Ross11,Huang14}. Simple theoretical methods are available for the computation of the generalized susceptibility for such reduced Hilbert space systems \cite{Becker77,Richards84,Yaouanc09,Yaouanc11}. An exhaustive theoretical study of that susceptibility for different types of fictitious spins 1/2 on a tetrahedron might be of great help for the interpretation of a large spectrum of data ranging from specific heat and ac susceptibility to neutron scattering and $\mu$SR.

\begin{acknowledgments}
This research project has been partially supported by the European Commission under the 
7th Framework Programme through the `Research Infrastructures' action of the
`Capacities' Programme, Contract No: CP-CSA\_INFRA-2008-1.1.1 Number
226507-NMI3. We thank STFC for the provision of muon beamtime at the ISIS Facility, UK. PDR thanks ISIS for financial support.
\end{acknowledgments}

\bibliography{reference.bib}

\end{document}